\documentclass[a4paper,10pt,twoside]{cpc-hepnp}

\usepackage{multicol}
\usepackage{graphicx}
\usepackage{booktabs}
\usepackage{amssymb,bm,mathrsfs,bbm,amscd}
\usepackage[tbtags]{amsmath}
\usepackage{lastpage}
\usepackage{CJK}

\begin{document}
\begin{CJK*}{GBK}{song}

\fancyhead[c]{\small Submitted to Chinese Physics C} \fancyfoot[C]{\small \thepage}

\footnotetext[0]{Received }

\title{Investigations on the charmless decays of $Y(4260)$\thanks{Supported by National Natural Science
Foundation of China (11275113,11205164) }}

\author{%
      LI Gang$^{1,3;1)}$\email{gli@mail.qfnu.edu.cn}%
\quad AN Chun-Sheng$^{2;2)}$\email{ancs@ihep.ac.cn}%
\quad LI Peng-Yu$^{1}$\\
\quad LIU Di$^{1}$
\quad ZHANG Xiao$^{1}$
\quad ZHOU Zhu$^{1}$}
\maketitle

\address{%
$^1$ Department of Physics, Qufu Normal University, Qufu
273165, China\\
$^2$Institute of High Energy Physics, and Theoretical Physics Center for Science Facilities,
Chinese Academy of Sciences, Beijing 100049, China\\
$^3$State Key Laboratory of Theoretical Physics, Institute of Theoretical Physics, Chinese Academy of Sciences, Beijing 100190, China
}

\begin{abstract}
Apart from the charmful decay channels of $Y(4260)$, the charmless decay channels of $Y(4260)$ also provide us a good platform to study the nature and the decay mechanism of $Y(4260)$.
In this paper, we propose to probe the structure of $Y(4260)$ through the charmless decays $Y(4260)\to VP$ via intermediate $D_1\bar D+c.c.$ meson loops, where $V$ and $P$ stand for light vector and pseudoscalar mesons, respectively. Under the molecule ansatz of $Y(4260)$, the predicted total branching ratio $BR_{VP}$ for all $Y(4260)\to VP$ processes are about
$(0.34^{+0.32}_{-0.23})\%$ to $(0.75^{+0.72}_{-0.52})\%$ with the cutoff parameter $\alpha=2\sim 3$. Numerical results show that the intermediate $D_1 \bar D+c.c.$ meson loops may be a possible transition mechanism in the $Y(4260)\to VP$ decays.
These predicted branching ratios are the same order to that of $Y(4260) \to Z_c^+(3900) \pi^-$, which may be an evidence of $D_1D$ molecule and  can be examined by the forthcoming BESIII data in the near future.
\end{abstract}



\footnotetext[0]{\hspace*{-3mm}\raisebox{0.3ex}{$\scriptstyle\copyright$}2013
Chinese Physical Society and the Institute of High Energy Physics
of the Chinese Academy of Sciences and the Institute
of Modern Physics of the Chinese Academy of Sciences and IOP Publishing Ltd}%

\begin{multicols}{2}

\section{Introduction}
\label{sec:introduction}

In the past decade, many new charmonium (or charmoniumlike),
i.e., the so-called $XYZ$ states have been observed experimentally,
which triggered a lot of theoretical
investigations on the
nature of exotic meson resonances beyond the conventional $q\bar{q}$ quark model~\cite{Brambilla:2010cs,Swanson:2006st,Eichten:2007qx,Voloshin:2007dx,Godfrey:2008nc,Drenska:2010kg}.
Among these observed $XYZ$ states, the resonance $Y(4260)$,
which was firstly observed by the BaBar Collaboration in the $\pi^+\pi^-
J/\psi$ invariant spectrum  in $e^+e^- \to \gamma_{ISR}\pi^+\pi^-
J/\psi$~\cite{Aubert:2005rm}, and then confirmed by both the
CLEO and Belle Collaborations~\cite{He:2006kg,Yuan:2007sj}, is a very interesting one
because of that its  mass $m=4263^{+8}_{-9}$ MeV~\cite{Olive:2014ss} is only about $30$-$40$ MeV below the $S$-wave $D_1 {\bar D} +c.c.$ threshold.
And very recently, the
new datum from BESIII confirms the signal in $Y(4260)\to
J/\psi\pi^+\pi^-$ with much higher
statistics~\cite{Ablikim:2013mio}. It indicates that it's worth to
study the structure and decays of $Y(4260)$.

Since the observation of $Y(4260)$, many different solutions were proposed to study the structure of $Y(4260)$.
These solutions include the $4S$ charmonium~\cite{LlanesEstrada:2005hz}, tetraquark $c{\bar c} s{\bar s}$ state~\cite{Maiani:2005pe}, charmonium
hybrid~\cite{Zhu:2005hp,Kou:2005gt,Close:2005iz}, $D_1\bar{D}$ molecule~\cite{Ding:2007rg,Ding:2008gr,Wang:2013cya},${^{\footnotemark[1]}}${\footnotetext[1]{There are two $D_1$ states of similar
masses, and the one in question should be the narrower one, i.e., the
$D_1(2420)$ ($\Gamma=27$ MeV), the $D_1(2430)$($\Gamma \simeq 384$
MeV) is too broad to form a molecular
state~\cite{Filin:2010se,Guo:2011dd,Guo:2013zbw}.}}
$\chi_{c1}\omega$ molecule~\cite{Yuan:2005dr},
$\chi_{c1}\rho$ molecule~\cite{Liu:2005ay}, hadrocharmonium state~\cite{Voloshin:2007dx,Dubynskiy:2008mq,Li:2013yka},  spin-triplet  $\Lambda_c$-${\bar \Lambda}_c$ baryonium states~\cite{Qiao:2005av,Qiao:2007ce,Chen:2011cta,Chen:2013sba}, a
cusp~\cite{vanBeveren:2009fb,vanBeveren:2009jk} or a non-resonance
explanation~\cite{vanBeveren:2010mg,Chen:2010nv} etc.
Under the $D_1 {\bar D}$ molecule ansatz,
some experimental observations can be described, such as the observation of $Z_c(3900)$ in
$e^+e^- \to \pi^+\pi^-
J/\psi$~\cite{Wang:2013cya}, the
production of $X(3872)$ in the $e^{+}e^{-}$ annihilation around the mass of
$Y(4260)$~\cite{Guo:2013zbw},
and the threshold behavior in the main decay
channels of $Y(4260)$~\cite{Liu:2013vfa} etc. In Ref.~\cite{Li:2013yka}, Li and Voloshin argue
that the hadrocharmonium interpretation of $Y(4260)$ may be more credible. Their argument is based on
the fact that the production of an $S$-wave pairs with $S_L^P=(3/2)^+$ and $S_L^P=(1/2)^-$ heavy mesons,
where $S_L$ is the sum of the spin of the light quark and the orbital angular momentum in the heavy mesons,
in $e^+e^-$ collisions is forbidden in the limit of exact
heavy quark spin symmetry. In Ref.~\cite{Li:2013yka}, it was also shown that both the rescattering
due to the process $D^* {\bar D}^* \to D_1{\bar D}$ and the mixing of the $D_1(2420)$ with the $D_1(2430)$ cannot
evade this suppressed production. They also considered the possible kinematic effects that might increase the amount
of the heavy quark spin symmetry (HQSS) violation and found that the kinematical effect is quite small at such energy. Thus, they concluded that the $S$-wave $D_1 {\bar D}$ production is suppressed. In Ref.~\cite{Wang:2013kra},  Wang {\it et al.} confront both the hadronic molecule and the hadrocharmonium interpretations of the $Y(4260)$ with the experimental data currently available. Although the production of $(3/2)^+$ and $(1/2)^-$ heavy meson pairs is suppressed in the heavy quark limit~\cite{Li:2013yka}, the
heavy quark spin symmetry breaking effects in the charm sector can be significant. So the resulting suppression for
the physical charm quark mass is not in conflict with the
interpretation that the main component of the
$Y(4260)$ is a $D_{1}\bar{D}$
molecule.

On the other hand, the intermediate meson loop transition as an important
nonperturbative dynamical mechanism has been extensively studied in the energy region of
charmonium~\cite{Li:1996yn,Cheng:2004ru,Anisovich:1995zu,Li:2011ssa,Li:2013jma,Li:2007ky,Wang:2012mf,Li:2012as,Li:2007xr,Achasov:1990gt,Achasov:1991qp,Achasov:1994vh,Achasov:2005qb,Zhao:2006dv,Zhao:2006cx,Wu:2007jh,Liu:2006dq,Zhao:2005ip,Li:2007au,Liu:2009dr,Li:2013zcr,Zhang:2009kr,Liu:2009vv,Liu:2010um,Wang:2012wj,Guo:2009wr,Guo:2010zk,Guo:2010ak}.
It is widely recognized that the intermediate meson loops may be closely related to some nonperturbative phenomena observed in
experiments~\cite{Achasov:1990gt,Achasov:1991qp,Achasov:1994vh,Achasov:2005qb,Wu:2007jh,Liu:2006dq,Cheng:2004ru,Anisovich:1995zu,Zhao:2005ip,Li:2007au,Liu:2009dr,Li:2013zcr,Zhang:2009kr,Liu:2009vv,Liu:2010um,Wang:2012wj,Guo:2009wr,Guo:2010zk,Guo:2010ak,Li:2013xia,Brambilla:2004wf,Brambilla:2004jw},
e.g. sizeable branching ratios for non-$D\bar D$ decay of
$\psi(3770)$~\cite{Achasov:1990gt,Achasov:1991qp,Achasov:1994vh,Achasov:2005qb,Liu:2009dr,Li:2013zcr,Zhang:2009kr},
the helicity selection rule violations in charmonium
decays~\cite{Liu:2009vv,Liu:2010um,Wang:2012wj}, isospin symmetry breaking in charmonium decays~\cite{Guo:2010zk,Guo:2010ak}.
Recently, this intermediate meson loops mechanism has been applied
to the production and decays of ordinary and exotic states~\cite{Wang:2013cya,Liu:2013vfa,Guo:2013zbw,Wang:2013hga,Cleven:2013mka,Wu:2013onz,Li:2013yla,Li:2014uia,Li:2014pfa}.

Recently, the charmful decay channels have been extensively used to constrain the reaction mechanism and gain  insights into the nature $Y(4260)$~\cite{Liu:2013vfa,Wang:2013hga,Cleven:2013mka,Wu:2013onz}.
Apart from the charmful decay channels of $Y(4260)$, the charmless decay channels of $Y(4260)$ are also a good platform to further study $Y(4260)$. In the present work, we study the charmless decays $Y(4260)\to VP$ via $D_1\bar{D}$ loop
with an effective Lagrangian approach (ELA) under the $D_1 \bar D + c.c.$ molecule ansatz. The paper is organized as follows. In
Sec.~2, we will briefly introduce the ELA and give
some relevant formulae, the numerical results are presented in Sec.~3, and
Sec.~4 contains a brief summary.

\section{The Model}
\label{sec:formula}

\begin{center}
\includegraphics[width=0.18\textwidth]{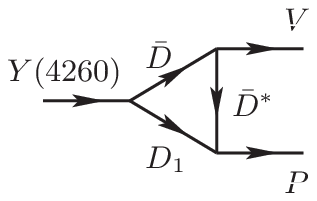}
\figcaption{\label{fig:feyn-fig1} The hadron-level diagrams for  $Y(4260) \to
VP$ with $D_1{\bar D}$ as the intermediate
states. $V$ and $P$ denote the light vector and pseudoscalar mesons, respectively.}
\end{center}
Generally speaking, all the possible intermediate meson exchange loops should be included in the calculation. In
reality, the breakdown of the local quark-hadron duality allows us to pick up the leading contributions as a reasonable approximation~\cite{Lipkin:1986bi,Lipkin:1986av}. For example, the intermediate states involving flavor changes turn out to be strongly suppressed. One reason is because of the large virtualities involved in the light meson loops. The other is because of the Okubo-Zweig-Iizuka-rule suppressions. In this work, we have assumed that $Y(4260)$ is dominated by the $S$-wave $ D_1{\bar D} +c.c.$ component and the  $D_1{\bar D} +c.c.$ mass threshold is only $30$ MeV above the $Y(4260)$, so we consider the S-wave $D_1{\bar D}$ meson loops as the leading contributions.

By assuming $Y(4260)$ is an S-wave $D_1\bar{D}$ molecular state, the
effective Lagrangian is constructed as
\begin{eqnarray}
\mathcal{L}_{Y(4260) D_1D}=i\frac {x} {{\sqrt 2}}(\bar{D}_a^\dag Y^\mu
D_{1a}^{\mu\dag}-\bar{D}_{1a}^{\mu\dag} Y^\mu D_a^\dag)+H.c., \label{eq:L-Y}
\end{eqnarray}
where $x$ is the coupling constant.

For a state slightly below an S-wave two-hadron threshold, the effective coupling constant of this state to the two-body channel, $g_{NR}$, is related to the probability of finding the two-hadron component in the physical wave function of the bound state, $c^2$, and the binding energy, $\epsilon=m_1+m_2-M$~\cite{Weinberg:1965zz, Baru:2003qq,Guo:2013zbw}
\begin{eqnarray}
g_{{\rm NR}}^2\equiv 16\pi (m_1+ m_2)^2  c^2\sqrt{\frac {2\epsilon}{\mu}} [1+ {\cal O}(\sqrt{2\mu\epsilon r})]\ , \label{eq:coupling-Y}
\end{eqnarray}
where $\mu=m_1 m_2/(m_1+m_2)$ is the reduced mass, and $r$ denotes the range of the forces. Notice that the coupling constant gets maximized for a pure bound state, which has $c^2 = 1$ by definition.

Using the masses of the $Y(4260)$, $D$ and $D_1$ given in PDG~\cite{Olive:2014ss}, we obtain the mass difference between the $Y(4260)$ and the $D_1 {\bar D} +c.c.$
threshold to be  $m_D+m_{D_1}-m_Y=27_{-8}^{+9}$ MeV.  Assuming that $Y(4260)$ is pure $DD_1$ molecule, which corresponding to hte probability of finding $D_1 \bar D$ component in the physical wave function of the bound states $c^2=1$, we obtain the coupling constant $x$
\begin{eqnarray} \label{eq:coupling-Y1}
|x|=14.62^{+1.11}_{-1.25}\pm 6.20 \ {\rm GeV} \, ,
\end{eqnarray}
where the first errors are due to the uncertainties of the binding
energies, and the second ones are from the the approximate nature
of Eq.~(\ref{eq:coupling-Y}).

The effective Lagrangian relevant to the light vector mesons can be obtained as
follows~\cite{Casalbuoni:1992gi,Casalbuoni:1992dx},
\begin{eqnarray}
{\cal L}_{{\cal V}}&=&ig_{{\cal D}^*{\cal D} {\cal V}}\epsilon_{\alpha\beta\mu\nu}
({\cal D} \stackrel{\leftrightarrow}{\partial_{\alpha}}{\cal D}^{*\beta\dagger}
-{\cal D}^{*\beta\dagger} \stackrel{\leftrightarrow}{\partial_{\alpha}}{\cal D}^{j})\partial^\mu {\cal V}^\nu \nonumber \\
&& +ig_{\overline{\cal D}^* {\overline
{\cal D}}{\cal V}}\epsilon_{\alpha\beta\mu\nu} (\overline{\cal D}
\stackrel{\leftrightarrow}{\partial_{\alpha}}\overline{\cal
D}^{*\beta\dagger} -\overline{\cal D}^{*\beta\dagger}
\stackrel{\leftrightarrow}{\partial_{\alpha}}\overline{\cal
D}^{j})\partial^\mu {\cal V}^\nu \nonumber \\
&& + H.c. \ , \label{eq:L-D*DV}
\end{eqnarray}
and the effective Lagrangian for the light pseudoscalar mesons are
constructed based on both heavy quark
spin-flavor transformation and chiral
transformation~\cite{Burdman:1992gh,Yan:1992gz,Casalbuoni:1996pg,Falk:1992cx}.
Accordingly, the interaction terms studied in the present work read
\begin{eqnarray}
\mathcal{L}_{{\cal P}} &=&
g_{D_1{\cal D}^*{\cal P}}[3D_{1}^\mu(\partial_\mu\partial_\nu {\cal P}){\cal D}^{*\dag
\nu}
-D_{1}^\mu(\partial^{\nu}\partial_\nu{\cal P}){\cal D}_{\mu}^{*\dag}] \nonumber\\
&&+ g_{{\bar D}_1{\bar {\cal D}}^*{\cal P}}[3\bar{{\cal D}}^{*\dag
\mu}(\partial_\mu\partial_\nu {\cal P})
\bar{D}_{1}^\nu-\bar{{\cal D}}^{*\dag \mu}(\partial^\nu \partial_\nu
{\cal P}) \bar{D}_{1\nu}]\nonumber \\
&&+H.c. \ , \label{eq:L-D1D*P}
\end{eqnarray}
with ${{{\cal D}}^{(*)}}=\left(D^{(*)+},D^{(*)0}, D^{(*)+}_s\right)$ and
${\bar {\cal D}^{(*)}}=\left(D^{(*)-},\bar{D}^{(*)0},D_s^{(*)-}\right)$.
${\cal P}$ and ${\cal V}$ denote the $3\times 3$ matrices for the pseudoscalar octet and vector nonet, respectively~\cite{Cheng:2004ru}, i.e.,

\end{multicols}

\begin{eqnarray}
\mathcal{P} =
\left(
  \begin{array}{ccc}
  \frac {\pi^0}{\sqrt{2}}+\frac {\eta\cos\alpha_P+\eta^\prime\sin\alpha_P} {\sqrt{2}} & \pi^+ & K^+ \cr
  \pi^- & -\frac {\pi^0} {\sqrt{2}}+\frac { \eta\cos\alpha_P +  \eta^\prime\sin\alpha_P} {\sqrt{2}} & K^0  \cr
K^- & {\bar K}^0 & - \eta\sin\alpha_P + \eta^\prime{\cos\alpha_P}
  \end{array}
\right),
\mathcal{V} =
\left(
\begin{array}{ccc} \frac {\rho^0} {\sqrt{2}}+\frac {\omega} {\sqrt{2}} & \rho^+ &
K^{*+} \cr
\rho^- & -\frac {\rho^0} {\sqrt{2}}+ \frac {\omega} {\sqrt{2}} & K^{*0}  \cr
K^{*-} & {\bar K}^{*0} & \phi
\end{array}
\right).
\end{eqnarray}

\begin{multicols}{2}
The physical states $\eta$ and $\eta^\prime$, which should be
linear combinations of $n{\bar n} = ({u\bar u} + {d\bar d})/\sqrt{2}$ and $s\bar{s}$,
are taken to be the following form
\begin{eqnarray}
|\eta\rangle &=& \cos\alpha_P|n\bar n\rangle  -\sin\alpha_P |s\bar s \rangle \, , \nonumber \\
|\eta^\prime \rangle &=& \sin\alpha_P |n\bar n\rangle  + \cos\alpha_P |s\bar s \rangle \, ,
\end{eqnarray}
where $\alpha_P \simeq \theta_P + \arctan {\sqrt{2}}$. Empirical value for the
pseudoscalar mixing angle $\theta_P$ should in a range of $-22^\circ
\sim -13^\circ$~\cite{Olive:2014ss}, and here we take $\theta_P = -19.3^\circ$~\cite{Liu:2006dq}.

And the coupling constants relevant to the light vector mesons in Eq.~(\ref{eq:L-D*DV}) read
\begin{eqnarray}
g_{{\cal D}^{*}{\cal D}{\cal V}}=-g_{{\overline{\cal
D}^{*}\overline{\cal D}}{\cal V}}=-\frac{1}{\sqrt{2}}\,\lambda g_{V}  \, ,
\end{eqnarray}
where $f_\pi$ = 132 MeV is the pion decay constant, and the
parameter $g_V$ is given by $g_V = {m_\rho /
f_\pi}$~\cite{Casalbuoni:1996pg}. By matching the form factor obtained from the light cone sum rule and that calculated from the Lattice QCD, we can obtain the parameter $\lambda=0.56$ GeV$^{-1}$~\cite{Isola:2003fh}.

In the chiral and heavy quark symmetry limit, the coupling constants relevant to the pseudoscalar mesons
in Eq.~(\ref{eq:L-D1D*P}) are
\begin{eqnarray}
g_{\rm
D^{*}D_1P}&=&g_{\rm\overline{D}^{*}\overline{D}_1P}=-\frac{\sqrt{6}}{3}\,\frac{h^\prime}{\Lambda_{\chi}f_{\pi}}\sqrt{\rm
m_{D^{*}}m_{D_1}} \ .
\end{eqnarray}
Here $\Lambda_\chi$ is the momentum scale characterising the convergence of the derivative expansion, usually taken as the chiral symmetry breaking scale $\Lambda_\chi \simeq 1$ GeV.
The coupling $h^\prime$, which is relevant to $\Delta_H$, i.e., the difference between
the charmed meson doublet mass and the mass of the heavy quark involved, can be obtained in a constituent
quark-meson model~\cite{Deandrea:1999pa}. If one take the value  $\Delta_H= 0.4\pm 0.1$ GeV, then one can obtain $h^\prime = 0.65^{+0.44}_{-0.30}$~\cite{Deandrea:1999pa}.
As the total $D_2^{* 0}$ width is dominated by the one pion mode in the chiral heavy meson Lagrangian, one can use the experimental
result of $49.0\pm 1.4$ MeV to extract an experimental value for $h^\prime$ to be $0.74\pm 0.01$~\cite{Olive:2014ss}. Here, we take $h^\prime = 0.74\pm 0.01$ as an estimate.

The loop transition amplitudes for the transitions in
Fig.~\ref{fig:feyn-fig1} can be expressed
in a general form in the effective Lagrangian approach as follows,
\begin{eqnarray}
{\cal A}_{fi}=\int \frac {d^4 q_2} {(2\pi)^4} \sum_{D^* \ \mbox{pol.}}
\frac {T_1T_2T_3} {a_1 a_2 a_3}{\cal F}(m_2,q_2^2)\, ,
\end{eqnarray}
where $T_i$ and $a_i = q_i^2-m_i^2 \ (i=1,2,3)$ are the vertex
functions and the denominators of the intermediate meson
propagators, respectively. As mentioned above, the mass of $Y(4260)$ is slightly below the S-wave $D_1\bar D$ threshold, so the off-shell effects of intermediate $D_1$ and ${\bar D}$ should be smaller than that of the exchanged particle. So  in order to take care of the off-shell effects of the exchanged particles~\cite{Li:1996yn,Locher:1993cc,Li:1996cj}, we adopt a monopole form factor
\begin{eqnarray}\label{ELA-form-factor}
{\cal F}(m_{2}, q_2^2) \equiv \frac
{\Lambda^2-m_{2}^2} {\Lambda^2-q_2^2} ,
\end{eqnarray}
with $\Lambda\equiv m_2+\alpha\Lambda_{\rm QCD}$, and the QCD energy
scale $\Lambda_{\rm QCD} = 220$ MeV.

\section{Numerical Results}
\label{sec:results}
\end{multicols}

\begin{center}
\tabcaption{\label{tab:br}The predicted branching ratios of $Y(4260)$ decays with
different $\alpha$ values. The uncertainties are dominated by the use
of Eq.~(\ref{eq:coupling-Y}).}
\footnotesize
\begin{tabular*}{170mm}{@{\extracolsep{\fill}}ccccccccc }
\hline
\toprule
Final states       & No Form factor & \multicolumn{2}{c} {Monopole Form Factor}  \\
&  & $\alpha=2.0$  & $\alpha=3.0$  \\
\hline
$\rho^0\pi^0$             & $(1.46^{+1.41}_{-1.01})\times 10^{-2}$   & $(8.93^{+8.58}_{-6.12})\times 10^{-4}$  & $(1.98^{+1.91}_{-1.36})\times 10^{-3}$\\
$\rho \pi$                & $(4.39^{+4.25}_{-3.03})\times 10^{-2}$   & $(2.61^{+2.51}_{-1.99})\times 10^{-3}$  & $(5.92^{+5.72}_{-4.11})\times 10^{-3}$\\
$K^{*+} K^- +c.c. $       & $(4.90^{+4.72}_{-3.37})\times 10^{-3}$   & $(1.09^{+1.06}_{-0.76})\times 10^{-4}$  & $(3.27^{+3.12}_{-2.25})\times 10^{-4}$\\
$K^{*0} {\bar K}^- +c.c.$ & $(4.96^{+4.78}_{-3.41})\times 10^{-3}$   & $(1.44^{+1.38}_{-0.99})\times 10^{-4}$  & $(3.21^{+3.09}_{-2.21})\times 10^{-4}$\\
$\omega \eta$             & $(1.37^{+1.33}_{-0.95})\times 10^{-2}$   & $(3.63^{+3.51}_{-2.51})\times 10^{-4}$  & $(8.18^{+7.88}_{-5.62})\times 10^{-4}$\\
$\omega \eta^\prime$      & $(1.25^{+1.21}_{-0.86})\times 10^{-2}$   & $(3.47^{+3.35}_{-2.39})\times 10^{-5}$  & $(8.38^{+8.13}_{-5.77})\times 10^{-5}$\\
$\rho\eta $               & $(2.93^{+2.83}_{-2.01})\times 10^{-7}$   & $(9.48^{+9.13}_{-6.52})\times 10^{-9}$  & $(1.96^{+1.89}_{-1.35})\times 10^{-8}$\\
$\rho\eta^\prime$         & $(8.18^{+7.88}_{-5.62})\times 10^{-7}$   & $(3.27^{+3.15}_{-2.25})\times 10^{-8}$  & $(6.52^{+6.27}_{-4.41})\times 10^{-8}$\\
$\omega\pi^0$             & $(5.22^{+5.02}_{-3.56})\times 10^{-7}$   & $(1.44^{+1.39}_{-1.00})\times 10^{-8}$  & $(3.09^{+2.97}_{-2.13})\times 10^{-8}$\\
Total                     & $(8.03^{+7.78}_{-5.52})\%$               & $(3.36^{+3.24}_{-2.31})\times 10^{-3}$  & $(7.48^{+7.22}_{-5.16})\times 10^{-3}$\\
\bottomrule
\end{tabular*}
\end{center}


\begin{multicols}{2}
\begin{center}
\includegraphics[width=0.4\textwidth]{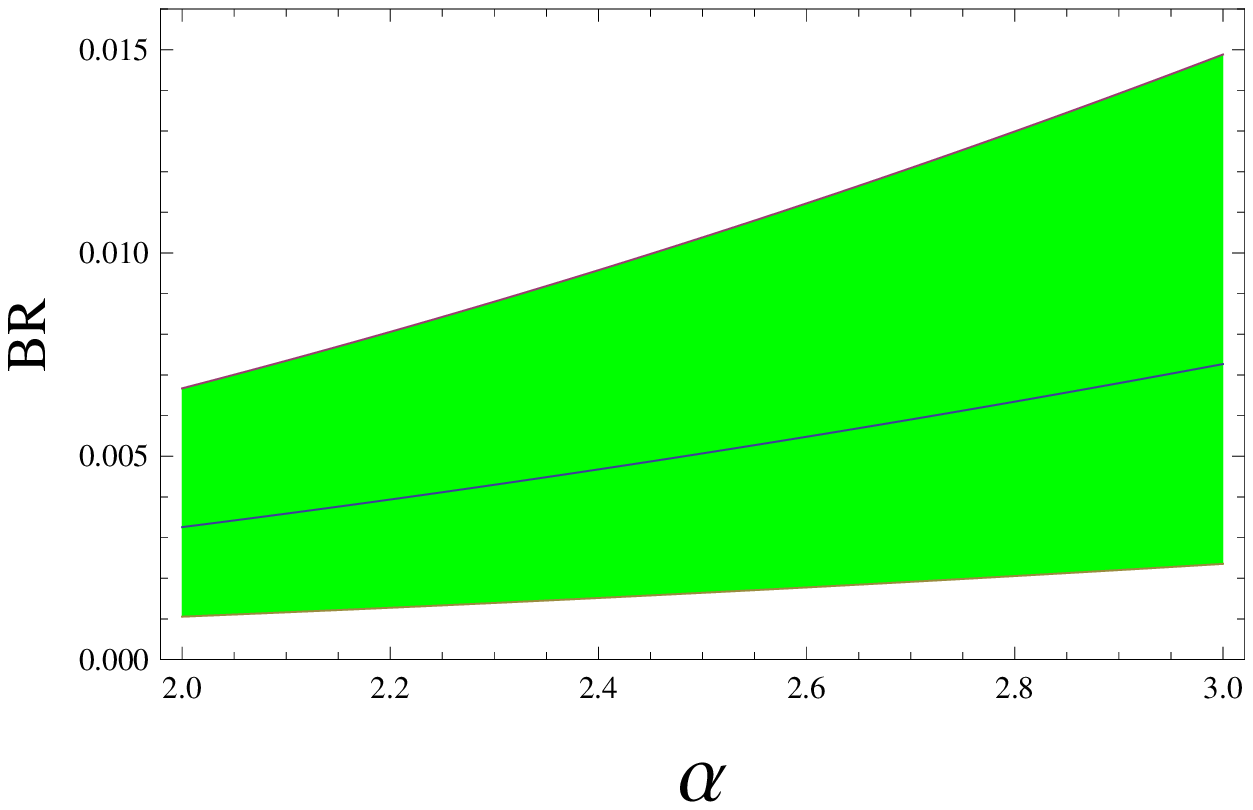}
\figcaption{\label{fig:br}The $\alpha$ dependence of the total branching ratios of $Y(4260)\to VP$. The upper and lower limits are obtained with the upper and lower limits of the coupling constant in Eq.~(\ref{eq:coupling-Y1}).}
\end{center}

\begin{center}
\includegraphics[width=0.4\textwidth]{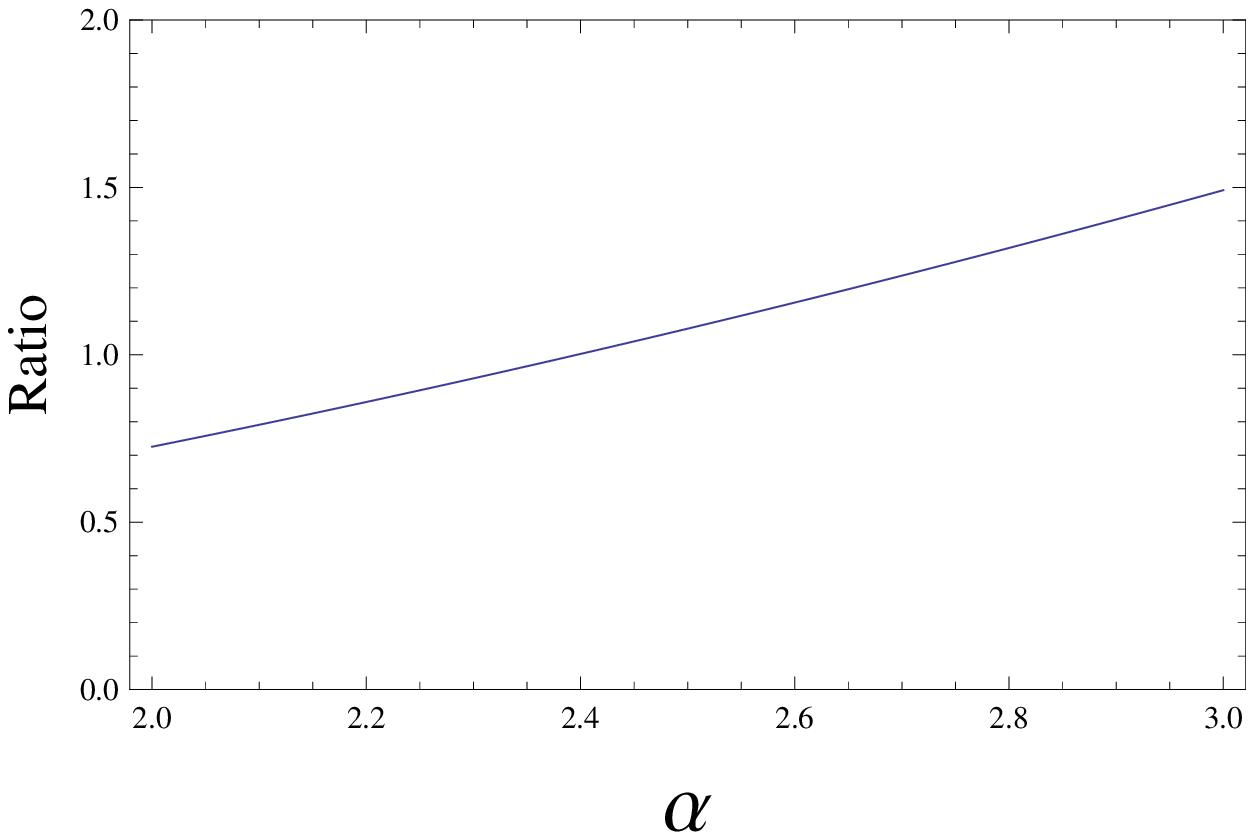}
\figcaption{\label{fig:ratio}The $\alpha$ dependence of the total branching ratios of $Y(4260)\to VP$. The upper and lower limits are obtained with the upper and lower limits of the coupling constant in Eq.~(\ref{eq:coupling-Y1}).}
\end{center}


The width of $Y(4260)$ is about $95\pm 14$ MeV~\cite{Olive:2014ss}, so we should take
into account the mass distribution of the $Y(4260)$ in the calculations of
its decay widths. Then the decay width of $Y(4260)\to VP$ can be calculated as
follow~\cite{Cleven:2011gp},
\begin{eqnarray}
\Gamma_{Y(4260)\to VP}&=&\frac {1} {W} \int_{(m_Y-2\Gamma_Y)^2}^{(m_Y+2\Gamma_Y)^2}ds \frac {(2\pi)^4}{2{\sqrt s}}\nonumber \\
&&\times  \int d\Phi_2 |{\cal A}|^2 \frac {1}{\pi} {\rm Im} (\frac {-1} {s-m_Y^2 +im_Y \Gamma_Y}), \nonumber \\
\end{eqnarray}
where $\cal A$ are the loop transition amplitudes for the processes in
Fig.~\ref{fig:feyn-fig1}. The factor $1/W$
with
\begin{eqnarray}
W= \frac {1}{\pi} \int_{(m_Y-2\Gamma_Y)^2}^{(m_Y+2\Gamma_Y)^2}
{\rm Im} (\frac {-1} {s-m_Y^2 +im_Y \Gamma_Y})ds
\end{eqnarray}
is used to normalize the spectral function of the
$Y(4260)$ state.

Before proceeding to the numerical results, we first discuss the possible uncertainties involved in the calculations. The first uncertainties is the assumption of the probability $c^2=1$ for the
$D_1 \bar{D}$ structure for $Y(4260)$. As shown in Eq.~(\ref{eq:coupling-Y}), the predicted branching ratios are proportional to probability $c^2$. The second one comes from the width effects of $Y(4260)$ and the final $\rho$ mesons. We have checked that the width effect of $\rho$ meson only causes a
minor change of about $1\%\sim 5\%$, which is because the mass of the final states are about $3$ GeV below $Y(4260)$.

In Fig.~\ref{fig:br}, we present the total branching ratio of all possible $Y(4260) \to VP$ in terms of the cutoff parameter $\alpha$. The upper and lower limits are obtained with the upper and lower limits of the coupling constant in~(\ref{eq:coupling-Y1}). As shown from this figure, there is no cusp structure in the curve. This is because the mass of $Y(4260)$ lies below
the intermediate $D_1 \bar D$ threshold. The branching ratios are not drastically sensitive to the cutoff
parameter, which indicates a reasonable cutoff of the ultraviolet
contributions by the empirical form factors to some extent.

To show the branching ratios of $Y(4260)$ to different
$VP$ channels explicitly, we list the predicted branching ratios of
$Y(4260)$ for each decay channel with $\alpha=2.0$ and $3.0$ in Table.~\ref{tab:br},
with comparison to the numerical results obtained without a form factor. Notice that
the given errors are from the
uncertainties of the the coupling constants in
Eq.~(\ref{eq:coupling-Y1}). As shown in Table~\ref{tab:br}, the total branching ratio of $Y(4260) \to VP$ is about $(8.03^{+7.78}_{-5.52})\%$ without form factor. Obviously, the obtained branching ratio in this way is somewhat larger
than expected. In principle,
since the $Y(4260)$ is taken to be a $D_1\bar D + c.c.$ molecule, so the main decay channel would
be $D^*{\bar D} \pi$.  It is because that the exchanged charmed mesons are usually off-shell, which indicates the necessity of considering the form factor.
As shown in the last two columns in Table~\ref{tab:br}, the total branching ratio of $Y(4260) \to VP$ are from $(3.36^{+3.24}_{-2.31})\times 10^{-3}$ to $(7.48^{+7.22}_{-5.16})\times 10^{-3}$
with the cutoff parameter $\alpha=2.0 \sim 3.0$.

For the isospin-violating channels, i.e., $Y(4260) \to \omega\pi^0$, $\rho\eta$, and $\rho\eta^\prime$,
the charged and neutral charmed meson loops would cancel out exactly in the isospin symmetry limit. In other words,
the mass difference between the $u$ and $d$ quark will lead to $m_{\cal D}^{(*)\pm} \neq m_{\cal D}^{(*)0}$ due to the isospin symmetry breaking. As a result, the charged and neutral charmed meson loops cannot completely cancel out, and the residue part will contribute to the isospin-violating amplitudes. The branching ratios of these isospin-violating channels as shown in Table~\ref{tab:br} are suppressed.
Differing from the isospin-violating channels, since there is no cancelations between the charged and neutral meson loops for the isospin isospin conserved channels, i.e., $Y(4260) \to \rho\pi$, $K^* {\bar K}+ c.c$, $\omega\eta$, and $\omega\eta^\prime$, so the calculated branching ratios of these channels are $3$-$4$
orders of magnitude larger than that of the isospin violated channels. As shown in this table, at the same $\alpha$,
the predicted branching ratios of $Y(4260)\to \omega\eta$ are one order larger than that of $Y(4260) \to \omega\eta^\prime$.
The reasons may attribute to the different $n{\bar n}$ component and different phase space. We suggest the experimental
measurements to test this point.

In order to better understand the decay mechanism of $Y(4260)$, we define the following ratio
\begin{eqnarray}\label{eq:ratio}
R=\frac {{\rm Br} (Y(4260)\to VP)} {{\rm Br} (Y(4260)\to Z_c^+(3900) \pi^-)} \, ,
\end{eqnarray}
which is plotted in Fig.~\ref{fig:ratio} for the dependence on the cutoff parameter. The ratio is less sensitive to the cutoff parameter, which is a consequence of the fact that the involved loops are the same. The predicted branching ratios for $Y(4260)\to VP$ are the same order to that of $Y(4260)\to Z_c(3900) \pi$. It may be an evidence for the molecule structure of $Y(4260)$  and can be tested by the experimental measurements in future.

\section{Summary}
\label{sec:summary}

In this work, we have investigated the charmless decays of
$Y(4260)$ in ELA, where $Y(4260)$ is considered as a $D_1 {\bar D}$ molecular state candidate. We explore the rescattering mechanism with the effective Lagrangian based on the heavy quark symmetry and chiral symmetry. The results show that the
$\alpha$ dependence of the branching ratios are not drastically
sensitive to some extent. With the commonly accepted $\alpha=2\sim 3$ range, we make a quantitative prediction for all $Y(4260)\to VP$ with $BR_{VP}$ from $(3.36^{+3.24}_{-2.31})\times 10^{-3}$ to $(7.48^{+7.22}_{-5.16})\times 10^{-3}$. These predicted branching ratios are the same order to that of $Y(4260) \to Z_c^+(3900) \pi^-$ with the molecular state assumption. It indicates that the intermediate $D_1 \bar D$ meson loops may be a possible mechanism in $Y(4260)\to VP$ decays.
Of course, the relevant calculations of these $Y(4260)\to VP$ channels in other models are also needed in order to study the nature of $Y(4260)$ deeply. We expect that with the help of precise measurements of various decay modes at BESIII, the nature of $Y(4260)$ and the decay mechanism of $Y(4260)\to VP$ can be investigated deeply. And the intermediate meson loops
mechanism can be established as a possible nonperturbative dynamics in the charmonium energy region, especially the initial states are close to the two particle thresholds.

\end{multicols}

\vspace{-1mm}
\centerline{\rule{80mm}{0.1pt}}
\vspace{2mm}

\begin{multicols}{2}

\end{multicols}

\clearpage

\end{CJK*}

\begin{thebibliography}{90}

\vspace{3mm}

\bibitem{Brambilla:2010cs}
  Brambilla~B, Eidelman S, Heltsley B K, Vogt R, Bodwin G T, Eichten E, Frawley A D and Meyer A B et al.
  Eur.\ Phys.\ J.\ C, 2011, {\bf 71}: 1534

\bibitem{Swanson:2006st}
  Swanson E S.
  Phys.\ Rept., 2006, {\bf 429}: 243

\bibitem{Eichten:2007qx}
  Eichten E, Godfrey S, Mahlke H and Rosner J H.
  Rev.\ Mod.\ Phys., 2008,  {\bf 80}: 1161

\bibitem{Voloshin:2007dx}
  Voloshin M B.
  Prog.\ Part.\ Nucl.\ Phys., 2008,  {\bf 61}: 455


\bibitem{Godfrey:2008nc}
  Godfrey S and Olsen S L.
  Ann.\ Rev.\ Nucl.\ Part.\ Sci., 2008,  {\bf 58}: 51

\bibitem{Drenska:2010kg}
  Drenska N, Faccini R, Piccinini F, Polosa F A, Renga F and Sabelli C,
  Riv.\ Nuovo Cim., 2010,  {\bf 033}: 633


\bibitem{Aubert:2005rm}
  Aubert B et al.  [BaBar Collaboration].
  Phys.\ Rev.\ Lett., 2005, {\bf 95}: 142001

\bibitem{He:2006kg}
  HE Q et al.  [CLEO Collaboration].
  Phys.\ Rev.\ D, 2006, {\bf 74}: 091104

\bibitem{Yuan:2007sj}
  YUAN C Z et al.  [Belle Collaboration].
  Phys.\ Rev.\ Lett., 2007,  {\bf 99}: 182004

\bibitem{Olive:2014ss}
  Olive K A et al.  [Particle Data Group Collaboration].
  Chin.\ Phys.\ C, 2014, {\bf 38}: 1

\bibitem{Ablikim:2013mio}
  Ablikim M et al.  [BESIII Collaboration].
  Phys.\ Rev.\ Lett., 2013,  {\bf 110}: 252001

\bibitem{LlanesEstrada:2005hz}
  Llanes-Estrada F J.
  Phys.\ Rev.\ D, 2005, {\bf 72}: 031503

\bibitem{Maiani:2005pe}
  Maiani L, Riquer V, Piccinini F and Polosa A D.
  Phys.\ Rev.\ D, 2005, {\bf 72}: 031502

\bibitem{Zhu:2005hp}
  ZHU S L.
  Phys.\ Lett.\ B, 2005, {\bf 625}: 212

\bibitem{Kou:2005gt}
  Kou E and Pene O.
  Phys.\ Lett.\ B, 2005, {\bf 631}: 164

\bibitem{Close:2005iz}
  Close F E and Page P R.
  Phys.\ Lett.\ B, 2005, {\bf 628}: 215

\bibitem{Ding:2007rg}
  DING G J, Zhu J J and YAN M L.
  Phys.\ Rev.\ D, 2008, {\bf 77}: 014033

\bibitem{Ding:2008gr}
  Ding G J.
  Phys.\ Rev.\ D, 2009, {\bf 79}: 014001

\bibitem{Wang:2013cya}
  WANG Q, Hanhart C and ZHAO Q.
  Phys.\ Rev.\ Lett., 2013,  {\bf 111}: 132003

\bibitem{Guo:2013zbw}
  GUO F K, Hanhart C, Mei{\ss}ner U G, WANG Q and ZHAO Q.
  Phys.\  Lett.\  B, 2013, {\bf 725}: 127

\bibitem{Filin:2010se}
  Filin A A, Romanov A, Baru V, Hanhart C, Kalashnikova Y X, Kudryavtsev A E, Mei{\ss}ner U G and Nefediev A V.
  Phys.\ Rev.\ Lett., 2010,  {\bf 105}: 019101

\bibitem{Guo:2011dd}
  GUO F K and Mei{\ss}ner U G.
  Phys.\ Rev.\ D, 2011, {\bf 84}: 014013

\bibitem{Yuan:2005dr}
  YUAN C Z, WANG P and MO X H.
  Phys.\ Lett.\ B, 2006, {\bf 634}: 399

\bibitem{Liu:2005ay}
  LIU X, ZENG X Q and LI X Q.
  Phys.\ Rev.\ D, 2005, {\bf 72}: 054023

\bibitem{Dubynskiy:2008mq}
  S.~Dubynskiy and Voloshin M B.
  Phys.\ Lett.\ B, 2008, {\bf 666}: 344

\bibitem{Li:2013yka}
  LI X and Voloshin M B.
  Phys.\ Rev.\  D, 2013, {\bf 588}: 034012

\bibitem{Qiao:2005av}
  QIAO C F.
  Phys.\ Lett.\ B, 2006, {\bf 639}: 263

\bibitem{Qiao:2007ce}
  QIAO C F.
  J.\ Phys.\ G, 2008, {\bf 35}: 075008

\bibitem{Chen:2011cta}
  CHEN Y D and QIAO C F.
  Phys.\ Rev.\ D, 2012, {\bf 85}: 034034

\bibitem{Chen:2013sba}
  CHEN Y D, QIAO C F, SHEN P. N and ZENG Z Q.
  Phys.\ Rev.\ D, 2013, {\bf 88}: 114007


\bibitem{vanBeveren:2009fb}
  Beveren E van and Rupp G.
  arXiv:0904.4351 [hep-ph]

\bibitem{vanBeveren:2009jk}
  Beveren E van  and Rupp G.
  Phys.\ Rev.\ D, 2009, {\bf 79}: 111501

\bibitem{vanBeveren:2010mg}
  Beveren E van, Rupp G and Segovia J.
  Phys.\ Rev.\ Lett., 2010,  {\bf 105} 102001

\bibitem{Chen:2010nv}
  CHEN D Y, HE J and LIU X.
  Phys.\ Rev.\ D, 2011, {\bf 83} 054021

\bibitem{Liu:2013vfa}
  LIU X H and LI G.
  Phys.\ Rev.\ D, 2013, {\bf 88}: 014013

\bibitem{Wang:2013kra}
  WANG Q, Cleven M, GUO F K, Hanhart C, Mei{\ss}ner U G, WU X G and ZHAO Q.
  Phys.\ Rev.\ D, 2014, {\bf 89}: 034001

\bibitem{Li:1996yn}
  LI X Q, Bugg D V and ZOU B S.
  Phys.\ Rev.\  D, 1997, {\bf 55}: 1421

\bibitem{Zhao:2006dv}
  ZHAO Q and ZOU B S.
  Phys.\ Rev.\  D, 2006, {\bf 74}: 114025

\bibitem{Zhao:2006cx}
  ZHAO Q.
  Phys.\ Lett.\  B, 2006, {\bf 636}: 197

\bibitem{Li:2011ssa}
  LI G and ZHAO Q.
  Phys.\ Rev.\ D, 2011, {\bf 84}: 074005

\bibitem{Li:2013jma}
  LI G, LIU X H and ZHAO Q.
  Eur.\ Phys.\ J.\ C, 2013, {\bf 73}: 2576

\bibitem{Li:2007ky}
  LI G, ZHAO Q and CHANG C H.
  J.\ Phys.\ G, 2008, {\bf 35}: 055002

\bibitem{Wang:2012mf}
  WANG Q, LI G and ZHAO Q.
  Phys.\ Rev.\ D, 2012, {\bf 85}: 074015

\bibitem{Li:2012as}
  LI G, SHAO F L, ZHAO C W and ZHAO Q.
  Phys.\ Rev.\ D, 2013, {\bf 87}: 034020

\bibitem{Li:2007xr}
  LI G and ZHAO Q.
  Phys.\ Lett.\ B, 2008, {\bf 670}: 55

\bibitem{Achasov:1990gt}
  Achasov N N and Kozhevnikov A A.
  Phys.\ Lett.\ B, 1991, {\bf 260}: 425

\bibitem{Achasov:1991qp}
  Achasov N N and Kozhevnikov A A.
  JETP Lett., 1991,  {\bf 54}: 193
  [Pisma Zh.\ Eksp.\ Teor.\ Fiz., 1991,  {\bf 54}: 197]

\bibitem{Achasov:1994vh}
  Achasov N N and Kozhevnikov A A.
  Phys.\ Rev.\ D, 1994, {\bf 49}: 275

\bibitem{Achasov:2005qb}
  Achasov N N and Kozhevnikov A A.
  Phys.\ Atom.\ Nucl., 2006,  {\bf 69}: 988

\bibitem{Zhang:2009kr}
  ZHANG Y J, LI G and ZHAO Q.
  Phys.\ Rev.\ Lett., 2009,  {\bf 102}: 172001

\bibitem{Liu:2009dr}
  LIU X, ZHANG BO and LI X Q.
  Phys.\ Lett.\  B, 2009, {\bf 675}: 441

\bibitem{Li:2013zcr}
  LI G, LIU X H, WANG Q and ZHAO Q.
  Phys.\ Rev.\ D, 2013, {\bf 88}: 014010

\bibitem{Wu:2007jh}
  WU J J, ZHAO Q and ZOU B S.
  Phys.\ Rev.\  D, 2007, {\bf 75}: 114012

\bibitem{Liu:2006dq}
  LIU X, ZENG X Q and LI X Q.
  Phys.\ Rev.\  D, 2006, {\bf 74}: 074003

\bibitem{Cheng:2004ru}
  CHENG H Y, CHUA C K and Soni A.
  Phys.\ Rev.\  D, 2005, {\bf 71}: 014030

\bibitem{Anisovich:1995zu}
  Anisovich V V, Bugg D V, Sarantsev A V and ZOU B S.
  Phys.\ Rev.\  D, 1995, {\bf 51}: 4619

\bibitem{Zhao:2005ip}
  ZHAO Q, ZOU B S and MA Z B.
  Phys.\ Lett.\  B, 2005, {\bf 631}: 22

\bibitem{Li:2007au}
  LI G, ZHAO Q and ZOU B S.
  Phys.\ Rev.\  D, 2008, {\bf 77}: 014010

\bibitem{Liu:2009vv}
  LIU X H and ZHAO Q.
  Phys.\ Rev.\  D, 2010, {\bf 81}: 014017

\bibitem{Wang:2012wj}
  WANG Q, LIU X H and ZHAO Q.
  Phys.\ Lett.\ B, 2012, {\bf 711}: 364

\bibitem{Liu:2010um}
  LIU X H and ZHAO Q.
  J.\ Phys.\ G, 2011, {\bf 38}: 035007

\bibitem{Guo:2009wr}
  GUO F K, Hanhart C and Mei{\ss}ner U G.
  Phys.\ Rev.\ Lett., 2009,  {\bf 103}: 082003  [Erratum-ibid., 2010,  {\bf 104}: 109901]

\bibitem{Guo:2010zk}
  GUO F K, Hanhart C, LI G, U.~G.~Mei{\ss}ner and ZHAO Q.
  Phys.\ Rev.\  D, 2010, {\bf 82}: 034025

\bibitem{Guo:2010ak}
  GUO F K, Hanhart C, LI G, U.~G.~Mei{\ss}ner and ZHAO Q.
  Phys.\ Rev.\  D, 2011, {\bf 83}: 034013

\bibitem{Li:2013xia}
  LI G,
  Eur.\ Phys.\ J.\ C, 2013, {\bf 73}: 2621


\bibitem{Brambilla:2004wf}
  N.~Brambilla et al.  [Quarkonium Working Group Collaboration].
  hep-ph/0412158

\bibitem{Brambilla:2004jw}
  Brambilla N, Pineda A, Soto J and Vairo A.
  Rev.\ Mod.\ Phys., 2005,  {\bf 77}: 1423

\bibitem{Wang:2013hga}
  WANG Q, Hanhart C and ZHAO Q.
  Phys.\ Lett.\ B, 2013, {\bf 725}: 106

\bibitem{Cleven:2013mka}
  Cleven M, WANG Q, GUO F K, Hanhart C, Mei{\ss}ner U G and ZHAO Q.
  arXiv:1310.2190 [hep-ph]

\bibitem{Wu:2013onz}
  WU X G, Hanhart C, WANG Q and ZHAO Q.
  Phys.\ Rev.\ D, 2014, {\bf 89}: 054038

\bibitem{Li:2013yla}
  LI G and LIU X H.
  Phys.\ Rev.\ D, 2013, {\bf 88}: 094008

\bibitem{Li:2014uia}
  LI G and WANG W.
  Phys.\ Lett.\ B, 2014, {\bf 733}: 100

\bibitem{Li:2014pfa}
  LI G, LIU X H and ZHOU Z.
  Phys.\ Rev.\  D, 2014, {\bf 90}: 054006

\bibitem{Lipkin:1986bi}
  Lipkin H J.
  Nucl.\ Phys.\ B, 1987, {\bf 291}: 720

\bibitem{Lipkin:1986av}
  Lipkin H J.
  Phys.\ Lett.\ B, 1986, {\bf 179}: 278

\bibitem{Weinberg:1965zz}
  Weinberg S.
  Phys.\ Rev., 1965,  {\bf 137} B672

\bibitem{Baru:2003qq}
  Baru V et al..
  Phys.\ Lett.\ B, 2004, {\bf 586}: 53

\bibitem{Casalbuoni:1992gi}
  Casalbuoni R, Deandrea A, Bartolomeo N Di, Gatto R, Feruglio F and Nardulli G.
  Phys.\ Lett.\ B, 1992, {\bf 292}: 371

\bibitem{Casalbuoni:1992dx}
  Casalbuoni R, Deandrea A, Bartolomeo N Di, Gatto R, Feruglio F and Nardulli G.
  Phys.\ Lett.\ B, 1993, {\bf 299}: 139

\bibitem{Burdman:1992gh}
  Burdman G and Donoghue J F.
  Phys.\ Lett.\ B, 1992, {\bf 280}: 287

\bibitem{Yan:1992gz}
  YAN T M, CHENG H Y, CHEUNG C Y, LIN G L, LIN C Y and YU H L.
  Phys.\ Rev.\ D, 1992,{\bf 46}: 1148 [Erratum-ibid.\ D, 1997, {\bf 55}: 5851]

\bibitem{Falk:1992cx}
  Falk A F and Luke M E.
  Phys.\ Lett.\ B, 1992, {\bf 292}: 119

\bibitem{Casalbuoni:1996pg}
  Casalbuoni R, Deandrea A, Bartolomeo N Di, Gatto R, Feruglio F and Nardulli G.
  Phys.\ Rept., 1997, {\bf 281}: 145

\bibitem{Isola:2003fh}
  Isola C, Ladisa M, Nardulli G and Santorelli P.
  Phys.\ Rev.\ D, 2003, {\bf 68}: 114001


\bibitem{Deandrea:1999pa}
  Deandrea A, Gatto R, Nardulli G and Polosa A D.
  JHEP, 1999, {\bf 9902}: 021



\bibitem{Locher:1993cc}
  Locher M P, LU Y and ZOU B S.
  Z.\ Phys.\ A, 1994, {\bf 347}: 281

\bibitem{Li:1996cj}
  LI X Q and ZOU B S.
  Phys.\ Lett.\ B, 1997, {\bf 399}: 297

\bibitem{Cleven:2011gp}
  Cleven M, GUO F K, Hanhart C and Mei{\ss}ner U G.
  Eur.\ Phys.\ J.\ A, 2011, {\bf 47}: 120

\end{thebibliography}
\end{document}